\documentclass[useAMS,usenatbib]{mn2e}

\usepackage{graphicx}

\title[]{A high precision chemical abundance analysis of the HAT-P-1 stellar binary: constraints on planet formation\thanks{The data presented herein were obtained at the W.M.\ Keck Observatory, which is operated as a scientific partnership among the California Institute of Technology, the University of California and the National Aeronautics and Space Administration. The Observatory was made possible by the generous financial support of the W.M.\ Keck Foundation.}}

\author[]{F. Liu,$^{1}$\thanks{E-mail: fan.liu@anu.edu.au}
M. Asplund,$^{1}$
I. Ram\'irez,$^{2}$
D. Yong,$^{1}$
J. Mel\'endez,$^{3}$\\
$^{1}$Research School of Astronomy and Astrophysics, Australian National University, Canberra, ACT 2611, Australia\\
$^{2}$McDonald Observatory and Department of Astronomy, University of Texas at Austin, 2515 Speedway, Austin, TX 78712-1205, USA\\
$^{3}$Departamento de Astronomia do IAG/USP, Universidade de Sao Paulo, Rua do Matao 1226, Sao Paulo 05508-900, SP, Brasil}

\begin{document}

\date{Accepted ? Received ?; in original form 2013 December 20}

\pagerange{\pageref{firstpage}--\pageref{lastpage}} \pubyear{2013}

\maketitle

\label{firstpage}

\begin{abstract}
We present a high-precision, differential elemental abundance analysis of the HAT-P-1 stellar binary based on high-resolution, high signal-to-noise ratio Keck/HIRES spectra. The secondary star in this double system is known to host a transiting giant planet while no planets have yet been detected around the primary star. 
The derived metallicities ([Fe/H]) of the primary and secondary stars are identical within the errors: $0.146 \pm 0.014$ dex ($\sigma$ = 0.033 dex) and $0.155 \pm 0.007$ dex ($\sigma$ = 0.023 dex), respectively. Extremely precise differential abundance ratios of 23 elements have been measured (mean error of $\sigma$([X/Fe]) = 0.013 dex) and are found to be indistinguishable between the two stars: $\Delta$[X/Fe] (secondary - primary) = +0.001 $\pm$ 0.006 dex ($\sigma$ = 0.008 dex). The striking similarity in the chemical composition of the two stellar components in HAT-P-1 is contrary to the possible 0.04 dex level difference seen in 16 Cyg A+B, which also hosts a giant planet, at least 3 times more massive than the one around HAT-P-1 secondary star. We conclude that the presence of giant planets does not necessarily imply differences in the chemical compositions of the host stars. The elemental abundances of each star in HAT-P-1 relative to the Sun show an identical, positive correlation with the condensation temperature of the elements; their abundance patterns are thus very similar to those observed in the majority of solar twins. In view of the \citet{m09}'s interpretation of the peculiar solar abundance pattern, we conclude that HAT-P-1 experienced less efficient formation of terrestrial planets than the Sun. This is in line with the expectation that the presence of close-in giant planets preventing the formation or survival of terrestrial planets.
\end{abstract}


\section{Introduction}

The components of binary systems are usually assumed to share the same origins and to have identical chemical compositions. Several studies, however, find that the elemental abundance differences in stellar binaries are in fact not uncommon \citep{g01,lg01,d04,d06,r11}. The sources of these abundance deviations remain unknown but one possible explanation is that any elemental abundance differences could relate to the processes of planet formation. 
It is well established that the dwarfs and sub-giants with higher metallicity have higher probability to form giant planets \citep{g97,fv05,us07,j10} yet the situation for the giants is still uncertain \citep{p07,h07,t08,m13}. Meanwhile whether planet formation can affect the chemical compositions of the host stars is less clear. \citet{m09} found small but statistically significant anomalies in the solar chemical composition compared to most solar twins and argued that these were due to the formation of terrestrial planets in the solar system that preferentially locked-up refractory elements (i.e. those easily condensing); the deficiency of refractory elements in the solar photosphere would disappear if some four Earth-masses of terrestrial planet material would be added into the present solar convective envelope \citep{c10}. \citet{m09} also found that the stars found {\em not} to have a close-in giant planet are more likely to resemble the Sun chemically, which suggests that the presence of close-in giant planets might prevent the formation of terrestrial planets. One could also speculate that in these systems, the smaller planets were accreted by the host stars during the migration process of Jupiter-like planets, thus removing the initial stellar abundance signature imprinted by the process of planet formation. \citet{l11} provide observational support for this idea based on Kepler data.

\citet{r11} demonstrated metallicity differences in the binary system 16 Cyg A+B to be $0.04 \pm 0.01$ dex (16 Cyg A is more metal-rich than 16 Cyg B) and related it to planet formation; 16 Cyg B is known to host a giant planet with a minimum mass of 1.68 Jupiter masses \citep{c97}. On the other hand, \citet{s11} found no such abundance differences in 16 Cyg A+B. The reasons for these contrary results remain unknown but given the possible connection between planet formation and stellar host composition, there is an urgent need for additional binary systems hosting planets to be exposed to a high precision abundance analysis. Here we present such a study for the HAT-P-1 stellar binary, in which a close-in giant planet orbits around the secondary star \citep{b07}; no planet detection around the primary star has been reported.

\section{Observations and data reduction}

We obtained high resolution (R = $\lambda/\Delta\lambda$ = 67,000), high signal-to-noise ratio (S/N $\simeq$ 300 per pixel) spectra of the HAT-P-1 stellar binary with the High Resolution Echelle Spectrometer (HIRES, \citealp{v94}) on the 10 m Keck I telescope on August 15, 2013. A solar spectrum with higher S/N ($\simeq$ 450 per pixel) was also obtained through observations of the asteroid Iris. The wavelength coverage of these spectra is nearly complete from 380 to 800 nm. The Keck-MAKEE pipeline was used for standard echelle spectra reduction which include bias subtraction, flat-fielding, scattered-light subtraction, spectral extraction and wavelength calibration. We normalized and co-added the spectra with IRAF\footnote{IRAF is distributed by the National Optical Astronomy Observatory, which is operated by Association of Universities for Research in Astronomy, Inc., under cooperative agreement with National Science Foundation.}.

\section{Stellar parameters and chemical abundance analysis}

We started the analysis by measuring the equivalent width (EW) for a number of lines. Our adopted line-list come mainly from solar abundance analysis of \citet{a09} but complemented with additional largely unblended lines from \citet{r03}, \citet{b05}, \citet{r07} and \citet{n09}; in a differential analysis such as ours the accuracy of the transition probabilities does not greatly influence the results. We measured the EW of each spectral line interactively using the \textit{splot} task in IRAF and discarded lines with equivalent width larger than 12 pm.
The final atomic-line data used for our abundance analysis are listed in Table A1. We performed a 1D, local thermodynamic equilibrium (LTE) abundance analysis with MOOG 2010 Version \citep{s73} using the ODFNEW grid of Kurucz model atmospheres \citep{c03}; in our differential analysis the choice of model atmospheres is inconsequential.

\begin{figure}
\centering
\includegraphics[width=\columnwidth]{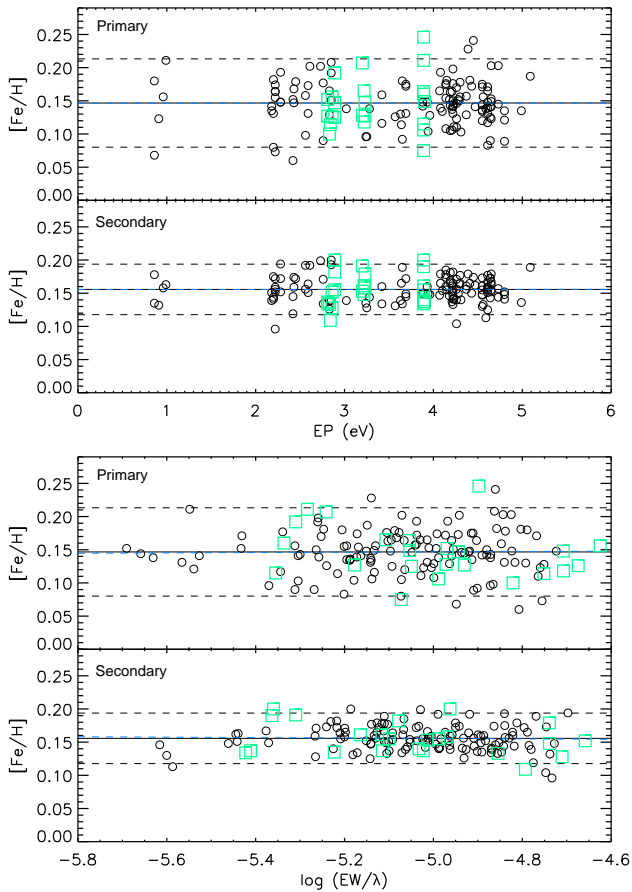}
\caption{Top panels: [Fe/H] of the HAT-P-1 stellar binary derived on a line-by-line basis with respect to the Sun as a function of excitation potential; open circles and green squares represent Fe\,{\sc i} and Fe\,{\sc ii} lines, respectively. Solid lines show the locations of mean [Fe/H], while dashed lines represent twice the standard deviation, $\pm 2 \sigma$. Bottom panels: same as in the top panels but as a function of reduced equivalent width.}
\label{f:parameters}
\end{figure}

The stellar parameters were derived using excitation and ionization balance of Fe\,{\sc i} and Fe\,{\sc ii} lines based on a line-by-line differential analysis relative to the Sun.
The adopted parameters for the Sun were $T_{\rm eff} = 5777$\,K, $\log g = 4.44$ [cgs], [Fe/H]=0.00 dex, $\varepsilon_{\rm t} = 1.00$\,km/s but we stress that the exact values are not crucial for our strictly differential study.
The stellar parameters for the two HAT-P-1 components were then established separately using a successively refined grid of stellar atmosphere models and the derived line-by-line differential abundances [Fe/H], finding the combination of $T_{\rm eff}$, $\log g$, [Fe/H] and $\varepsilon_{\rm t}$ that minimized the slopes in [Fe\,{\sc i}/H] versus excitation potential and reduced equivalent width as well as the difference between [Fe\,{\sc i}/H] and [Fe\,{\sc ii}/H]. We required the derived average [Fe/H] to be within 0.005 dex of the value used in the model atmosphere. This iterative procedure was considered converged when the grid step-size was $\Delta T_{\rm eff} = 1$ K, $\Delta \log g = 0.01$ and $\Delta \varepsilon_{\rm t} = 0.01$ km/s. No sigma clipping was implemented in this work. The final adopted stellar parameters are listed in Table \ref{t:parameters}, which satisfy the excitation and ionization balance in a differential sense (Fig. \ref{f:parameters}).

The uncertainties in the stellar parameters were calculated based on the procedure laid out by \citet{e10} (see also \citet{b14}), which accounts for the co-variances between changes in the stellar parameters and the differential abundances. Table \ref{t:parameters} lists the inferred errors, which highlights the excellent precision achieved:  $\sigma T_{\rm eff} = 17$ and 8\,K, respectively. These extremely low values for the errors correspond to the internal uncertainties of the differential method. Comparisons between sets of parameters derived in different studies show that the external uncertainties are usually higher \citep{b14}. Our analysis demonstrates that the primary star is 200\,K hotter than the secondary star while the metallicities of the primary and secondary stars are indistinguishable within the uncertainties: [Fe/H]=$0.146 \pm 0.014$ dex ($\sigma$ = 0.033 dex) and $0.155 \pm 0.007$ dex ($\sigma$ = 0.023 dex), respectively. Here the uncertainties were derived using the Epstein approach while the values of $\sigma$ represent the standard deviations of [Fe/H].

\begin{table} 
\caption{Stellar atmospheric parameters for HAT-P-1}
\label{t:parameters}
\begin{tabular}{@{}lccc@{}}
\hline
Parameter & Primary & Secondary & S - P\\
$T_{\rm eff}$ (K) & 6251 $\pm$ 17 & 6049 $\pm$ 8 & $-$202 $\pm$ 11\\
$\log g$ (cgs) & 4.36 $\pm$ 0.03 & 4.43 $\pm$ 0.02 & $+$0.07 $\pm$ 0.03\\
$\rm {[Fe/H]}$ (dex) & 0.146 $\pm$ 0.014 & 0.155 $\pm$ 0.007 & $+$0.009 $\pm$ 0.009\\
$\varepsilon_{\rm t}$ (km/s) & 1.45 $\pm$ 0.03 & 1.22 $\pm$ 0.02 & $-$0.23 $\pm$ 0.02\\
\hline
\end{tabular}
\end{table}

\begin{table}
\caption{Differential elemental abundances for HAT-P-1}
\label{t:abundances}
\begin{tabular}{@{}lrrr@{}}
\hline
Element & Primary$^a$ & Secondary$^a$ & $\Delta$[X/Fe]$^b$ \\
$ $[C\,{\sc i}/Fe]   & $-$0.158 $\pm$ 0.036 & $-$0.156 $\pm$ 0.030 &  0.002 $\pm$ 0.015 \\
$ $[O\,{\sc i}/Fe]   & $-$0.063 $\pm$ 0.024 & $-$0.067 $\pm$ 0.034 & $-$0.004 $\pm$ 0.020 \\
$ $[Na\,{\sc i}/Fe] & $-$0.067 $\pm$ 0.021 & $-$0.065 $\pm$ 0.008 &  0.002 $\pm$ 0.016 \\
$ $[Mg\,{\sc i}/Fe] & $-$0.060 $\pm$ 0.020 & $-$0.050 $\pm$ 0.008 &  0.011 $\pm$ 0.014 \\
$ $[Al\,{\sc i}/Fe]  & $-$0.025 $\pm$ 0.018 & $-$0.019 $\pm$ 0.011 &  0.006 $\pm$ 0.020 \\
$ $[Si\,{\sc i}/Fe]  & $-$0.004 $\pm$ 0.012 & $-$0.002 $\pm$ 0.007 &  0.001 $\pm$ 0.008 \\
$ $[S\,{\sc i}/Fe]   & $-$0.090 $\pm$ 0.021 & $-$0.097 $\pm$ 0.015 & $-$0.007 $\pm$ 0.011 \\
$ $[Ca\,{\sc i}/Fe]  &  0.009 $\pm$ 0.011 &  0.009 $\pm$ 0.008 &  0.000 $\pm$ 0.009 \\
$ $[Sc\,{\sc ii}/Fe] &  0.052 $\pm$ 0.017 &  0.042 $\pm$ 0.014 & $-$0.010 $\pm$ 0.012 \\
$ $[Ti\,{\sc i}/Fe]  &  0.006 $\pm$ 0.012 &  0.006 $\pm$ 0.008 &  0.000 $\pm$ 0.009 \\
$ $[Ti\,{\sc ii}/Fe] &  0.031 $\pm$ 0.015 &  0.023 $\pm$ 0.010 & $-$0.008 $\pm$ 0.012 \\
$ $[V\,{\sc i}/Fe]   &  0.017 $\pm$ 0.019 &  0.014 $\pm$ 0.012 & $-$0.003 $\pm$ 0.014 \\
$ $[Cr\,{\sc i}/Fe]  & $-$0.032 $\pm$ 0.011 & $-$0.018 $\pm$ 0.008 &  0.014 $\pm$ 0.009 \\
$ $[Cr\,{\sc ii}/Fe] & $-$0.032 $\pm$ 0.018 & $-$0.022 $\pm$ 0.015 &  0.010 $\pm$ 0.013 \\
$ $[Mn\,{\sc i}/Fe] & $-$0.055 $\pm$ 0.018 & $-$0.044 $\pm$ 0.018 &  0.011 $\pm$ 0.012 \\
$ $[Co\,{\sc i}/Fe] & $-$0.031 $\pm$ 0.015 & $-$0.023 $\pm$ 0.011 &  0.008 $\pm$ 0.014 \\
$ $[Ni\,{\sc i}/Fe]  & $-$0.017 $\pm$ 0.011 & $-$0.008 $\pm$ 0.006 &  0.009 $\pm$ 0.008 \\
$ $[Cu\,{\sc i}/Fe] & $-$0.094 $\pm$ 0.012 & $-$0.102 $\pm$ 0.015 & $-$0.008 $\pm$ 0.010 \\
$ $[Zn\,{\sc i}/Fe] & $-$0.112 $\pm$ 0.027 & $-$0.106 $\pm$ 0.021 &  0.007 $\pm$ 0.009 \\
$ $[Sr\,{\sc i}/Fe]  &  0.043 $\pm$ 0.019 &  0.031 $\pm$ 0.015 & $-$0.012 $\pm$ 0.014 \\
$ $[Sr\,{\sc ii}/Fe] & $-$0.016 $\pm$ 0.019 & $-$0.018 $\pm$ 0.015 & $-$0.002 $\pm$ 0.014 \\
$ $[Y\,{\sc ii}/Fe]   &  0.019 $\pm$ 0.042 &  0.023 $\pm$ 0.029 &  0.004 $\pm$ 0.017 \\
$ $[Zr\,{\sc ii}/Fe]  &  0.021 $\pm$ 0.013 &  0.036 $\pm$ 0.012 &  0.015 $\pm$ 0.012 \\
$ $[Ba\,{\sc ii}/Fe] &  0.072 $\pm$ 0.029 &  0.056 $\pm$ 0.013 & $-$0.016 $\pm$ 0.018 \\
$ $[La\,{\sc ii}/Fe] &  0.058 $\pm$ 0.017 &  0.065 $\pm$ 0.009 &  0.008 $\pm$ 0.014 \\
$ $[Ce\,{\sc ii}/Fe] &  0.030 $\pm$ 0.027 &  0.028 $\pm$ 0.023 & $-$0.002 $\pm$ 0.015 \\
\hline
\end{tabular}
$^a$ Relative to the Sun \\
$^b$ Secondary star relative to primary star
\end{table}

Having established the stellar parameters for the binary components, we derived chemical abundances for 23 elements: C, O, Na, Mg, Al, Si, S, Ca, Sc, Ti, V, Cr, Mn, Co, Ni, Cu, Zn, Sr, Y, Zr, Ba, La and Ce. Three elemental abundances (Sr, Zr and Ba) were derived through spectrum synthesis\footnote{The synthetic spectra were convolved with a Gaussian representing the combined effect of the atmospheric turbulence, stellar rotation, and the instrumental profile. The values of the broadening parameters for HAT-P-1 primary and secondary stars were 9.0, 6.0 km/s, respectively.}. Hyperfine structure splitting was considered for Sc, V, Cr, Mn, Cu and Ba \citep{k95}. Departures from LTE were considered for oxygen according to \citet{r07}. We also compared the NLTE corrections with those from \citet{f09} and note that the difference between two studies is 0.09 dex for the primary star and 0.06 dex for the secondary star, while the difference is 0.03 dex when comparing the primary and secondary star differentially.

The strictly line-by-line differential analysis greatly reduces the errors from atomic data and shortcomings in the 1D LTE modeling of the stellar atmospheres and spectral line formation. The abundances were determined using both the Sun and the primary star as reference stars; the inferred chemical compositions and associated $1\sigma$ uncertainties are listed in Table \ref{t:abundances}. The errors in the differential abundances correspond to the standard error of the mean added in quadrature to the errors introduced by the uncertainties in our atmospheric parameters following the method of \citet{e10}. All elemental abundances (with the primary as reference star) have uncertainties $\le 0.020$\,dex, which further underscores the advantages with a strictly differential analysis. The mean abundance difference of all elements between the secondary and primary is +0.001 $\pm$ 0.006 dex ($\sigma$ = 0.008 dex, secondary - primary), with no elemental abundance differing by more than 0.016\,dex (18 out of 24 elements differ by $\le 0.010$\,dex). Indeed, of the 26 species in Table \ref{t:abundances}, only three have differences outside the $1\sigma$ errors, which may suggest that we have in fact been overly conservative in the uncertainty estimations. For all purposes, the primary and secondary star in HAT-1-P are chemically indistinguishable. 

\section{Discussion}

\begin{figure}
\includegraphics[width=\columnwidth]{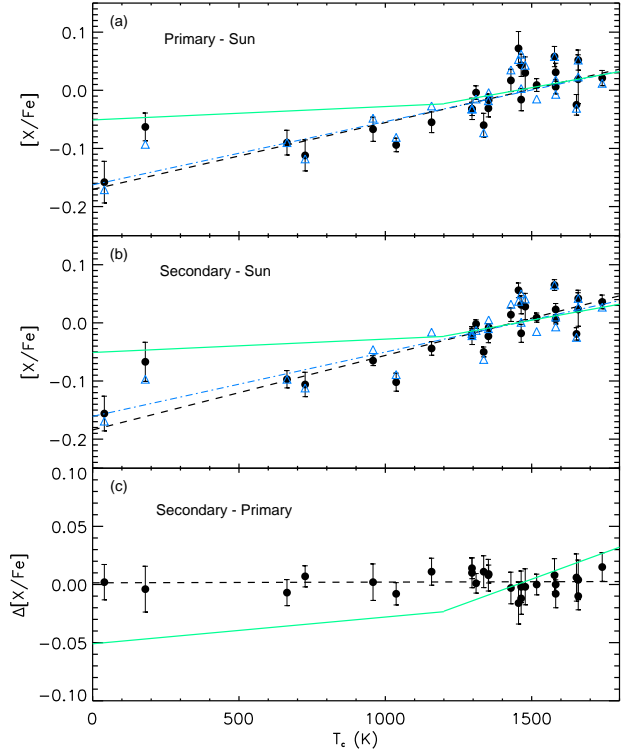}
\caption{Differential elemental abundances of HAT-P-1 stellar binary relative to our solar abundances and to each other as a function of dust condensation temperature; filled circles and blue triangles represent [X/Fe] without and with GCE corrections, respectively. Black dashed lines and blue dot-dashed lines show the fitting slopes of our results without and with GCE corrections, respectively. Green solid lines show the mean trend of solar twin stars according to \citet{m09}.}
\label{f:tcondensation}
\end{figure}

The focus of this discussion is to examine the possible connection between planet formation and stellar host composition in the HAT-P-1 binary. Given that the secondary star in the HAT-P-1 stellar binary is known to harbor a giant planet, our high precision chemical abundances may place new constraints on the planet formation process, at least in this system.

As noted in the introduction, \citet{m09} and follow-up studies \citep{r09,r10} discovered that the Sun shows deficiency in refractory elements relative to volatiles when compare to the majority ($\sim 80-90$\%) of solar twins. The deficiencies correlate with the condensation temperature ($T_{\rm c}$) of the elements such that the abundances of refractory elements ($T_{\rm c} \geq$ 900 K) decrease (Sun - solar twins) with increasing $T_{\rm c}$. They argue that the special abundance pattern of the Sun is due to dust condensation and terrestrial planet formation in the proto-solar disk that for some reason proceeded more efficiently than for the majority of solar twins. They then argue that terrestrial planets over giant planets as the cause for the peculiar abundance signature due to the presence of a break at $T_{\rm c} \approx 1200$\,K (much higher than the expected temperatures in the proto-planetary disk where the solar system giant planets formed), the required amount of refractory material necessary to imprint the signature ($4$\,M$_\oplus$) and the higher frequency of stars sharing the solar abundance pattern that do {\em not} have a close-in giant planet. 

We note however that \citet{o14} propose that the abundance differences found by \citet{m09} are not the result of planet formation but are imprinted by dust-gas separation in the interstellar medium prior to star formation based on their finding that all of their 14 stars in the open cluster M67 resemble more the Sun than the solar twins of \citet{m09}. They conclude that the Sun formed in an unusually dense stellar environment like M67. The existence of a high temperature break in $T_{\rm c}$ and the apparent correlation with absence of close-in giant planets are not easily understood in that scenario however. 

Our high quality data allow us to make robust conclusions about the [X/Fe] -- $T_{\rm c}$ slopes of the HAT-P-1 stellar binary. Fig. \ref{f:tcondensation} shows the differential abundances of HAT-P-1 primary and secondary star relative to the Sun and relative to each other versus $T_{\rm c}$ \citep{l03}\footnote{[O/Fe] of both stars relative to the Sun would fall to the fitting trends while $\Delta$[O/Fe] (relative to each other) would be 0.03 dex larger if NLTE corrections are adopted from \citet{f09}.}. All the elemental abundances were used to derive the slopes. The slopes of linear fitting for both stars compared to the Sun are positive and identical within errors: $(1.15 \pm 0.10) \times 10^{-4}$ dex K$^{-1}$ and $(1.28 \pm 0.08) \times 10^{-4}$ dex K$^{-1}$ for primary and secondary star, respectively. These slopes are very similar to the trends of refractories of the average of solar twins relative to the Sun \citep{m09,r09}, which with their interpretation would imply that both binary components formed less terrestrial planets than the Sun. This is consistent with the expectation that the presence of close-in giant planets prevents the formation or survival of terrestrial planets \citep{il04}. The positive slopes can also arise from Galactic chemical evolution (GCE). Therefore we applied the GCE corrections on our [X/Fe] values based on the studies of \citet{gh13}. We adopted the \citet{gh13}'s data and fitting trends to derive the values of [X/Fe] at [Fe/H] $\sim$ 0.15 dex to correct our results. The final results with GCE corrections only show tiny differences of the general trends (see Fig. \ref{f:tcondensation}a,b) which indicate that these positive slopes can not be erased even after GCE corrections.

When comparing the two HAT-P-1 components relative to each other, the slope of [X/Fe] -- $T_{\rm c}$ is non-existent (Fig. \ref{f:tcondensation}c): $(0.60 \pm 6.36) \times 10^{-6}$ dex K$^{-1}$. As stated before, the mean elemental abundance difference between the secondary and primary star is +0.001 $\pm$ 0.006 dex ($\sigma$ = 0.008 dex, secondary - primary). Clearly, the two stars have indistinguishable chemical compositions, which is interesting given the detection of a close-in giant planet with mass $\sim 0.53\,{\rm M_{Jupiter}}$ around the secondary star. We conclude that the formation process of giant planets does not necessarily affect the chemical pattern of the host star, which supports the conclusions of \citet{m09} and \citet{r09}. This is contrary to the difference of $0.04 \pm 0.01$\,dex seen in 16 Cyg A+B \citep{r11} but is consistent with the results from \citet{s11}.

Assuming for the moment that the 16 Cyg abundance differences are real, one possible explanation could be the higher mass ($2.4\,{\rm M_{Jupiter}}$) of the 16 Cyg planet \citep{p13}: it is still possible that such a more massive planet imprints a chemical signature in the host star. We note, however, the stellar masses in HAT-P-1 are slightly higher ($1.16$ and $1.12$\,$M_{\sun}$, \citealp{b07}) than in 16 Cyg ($1.05$ and $1.00$\,$M_{\sun}$, \citealp{r11}), which makes the convection zone less massive and thus more prone to chemical imprints from planet formation. Albeit the smaller convection zone in HAT-P-1 would make it easier to imprint a planet signature, higher mass stars seem to have shorter disk lifetimes \citep{w11}, making thus more difficult to imprint any planet signature in HAT-P-1 than in 16 Cyg. Which of these effects that dominate would depend on the exact size of the convection zone at the time of the accretion and the amount of material heavier than Helium locked up in the giant planet. For the time being, we conclude that the formation of giant planets do not necessarily have to introduce chemical signatures in their host stars.

Our detailed study of the HAT-P-1 double system underscores how high precision differential abundance measurements in binary stars with planets can provide important constraints on planet formation. Further efforts are needed to examine the physical characteristics and chemical abundances for additional stellar binaries with giant or terrestrial planets in order to understand the formation and evolution of planetary systems.

\section*{Acknowledgements}

This work has been supported by the Australian Research Council (grants FL110100012 and DP120100991). J. M. thanks FAPESP (2012/24392-2). The authors thank the ANU Time Allocation Committee for awarding observation time to this project. The authors wish to recognize and acknowledge the very significant cultural role and reverence that the summit of Mauna Kea has always had within the indigenous Hawaiian community. We are most fortunate to have the opportunity to conduct observations from this mountain.

\section*{SUPPLEMENTARY MATERIAL}

\noindent\\
Table A1. Atomic-line data used for our abundance analysis.

\begin{tabular}{@{}lccr@{}}
\hline
Wavelength (\AA) & Element & EP (eV) & $\log gf$ \\
5052.167 &  C \sc{i}  &  7.68  &  $-$1.30 \\
7113.179 &  C \sc{i}  &  8.65  &  $-$0.76 \\
7116.960 &  C \sc{i}  &  8.65  &  $-$0.91 \\
7771.944 &  O \sc{i}  &  9.15  &  0.35 \\
7774.161 &  O \sc{i}  &  9.15  &  0.22 \\
7775.390 &  O \sc{i}  &  9.15  &  0.00 \\
\hline
\end{tabular}

\noindent
Note. --- The entire table is available for this article online. A portion is shown here for guidance regarding its content.

\bsp

\label{lastpage}

\end{document}